\documentclass[iop]{emulateapj}  
\usepackage{graphicx}  

\def\simlt{\mathrel{\rlap{\lower 3pt\hbox{$\sim$}}\raise 2.0pt\hbox{$<$}}}
\def\simgt{\mathrel{\rlap{\lower 3pt\hbox{$\sim$}} \raise 2.0pt\hbox{$>$}}}
\def\lsim{\mathrel{\rlap{\lower 3pt\hbox{$\sim$}}\raise 2.0pt\hbox{$<$}}}
\def\gsim{\mathrel{\rlap{\lower 3pt\hbox{$\sim$}} \raise 2.0pt\hbox{$>$}}}

\def\Zsun{{\rm Z}_{\odot}}

\shortauthors{Salvaterra et al.}
\shorttitle{The GRB Luminosity Function}

\begin{document}
\title{A complete sample of bright {\it Swift} Long Gamma-Ray Bursts I: Sample
presentation, Luminosity Function and evolution}
\author{
R.~Salvaterra,\altaffilmark{1}
S.~Campana\altaffilmark{2}, 
S.D.~Vergani\altaffilmark{2,3}
S.~Covino\altaffilmark{2}
P.~D'Avanzo\altaffilmark{2}
D.~Fugazza\altaffilmark{2} 
G.~Ghirlanda\altaffilmark{2}
G,~Ghisellini\altaffilmark{2} 
A.~Melandri\altaffilmark{2}
L.~Nava\altaffilmark{4} 
B.~Sbarufatti\altaffilmark{2}
H.~Flores\altaffilmark{3}
S.~Piranomonte\altaffilmark{5}
G.~Tagliaferri\altaffilmark{2}
}
\altaffiltext{1}{INAF, IASF Milano, via E. Bassini 15, I-20133 Milano,
  Italy, ruben@lambrate.inaf.it}
\altaffiltext{2}{INAF, Osservatorio Astronomico di Brera, via E. Bianchi 46, I-23807 Merate
(LC), Italy}
\altaffiltext{3}{Laboratoire GEPI, Observatoire de Paris, CNRS-UMR8111, Univ. Paris-Diderot 5 place Jules Janssen, 92195 Meudon, France}
\altaffiltext{4}{SISSA, via Bonomea 265, I-34136 Trieste, Italy}
\altaffiltext{5}{INAF, Osservatorio Astronomico di Roma, Via Frascati 33, 00040, Monte Porzio Catone, Rome, Italy}

\begin{abstract}
We present a carefully selected sub-sample of {\it Swift} Long Gamma-ray Bursts 
(GRBs), that is complete in redshift. The sample is constructed by considering 
only bursts with favorable observing conditions for ground-based follow-up 
searches, that are bright in the 15-150 keV {\it Swift}/BAT band, i.e. 
with 1-s peak photon fluxes in excess to 2.6 ph s$^{-1}$ cm$^{-2}$. 
The sample is composed by 58 bursts, 52 of them with redshift for  a 
completeness level of $90$\%, while another two have a redshift constraint,
reaching  a completeness level of 95\%.
For only three bursts we have no constraint on the redshift.
The high level of redshift completeness allows us for
the first time to constrain the GRB luminosity function and its evolution 
with cosmic times in a unbiased way.
We find that strong evolution in luminosity ($\delta_l=2.3\pm 0.6$) or in 
density ($\delta_d=1.7\pm 0.5$) is required in order to account for the 
observations. The derived redshift distribution in the two scenarios
are consistent with each other, in spite of their different intrinsic redshift distribution. 
This calls for other indicators to distinguish among different evolution models.
Complete samples are at the base of any population studies.
In future works we will use this unique sample of {\it Swift} bright GRBs
to study the properties of the  population of long GRBs.
\end{abstract}


\section{Introduction}

Gamma-ray bursts are powerful flashes of high--energy photons occurring
at an average rate of a few per day throughout the Universe.
They are detected at all redshifts, from the
local Universe up to the extreme high redshifts (Salvaterra et al. 2009;
Tanvir et al. 2009; Cucchiara et al. 2011a). Our knowledge of the
distribution of long GRBs through cosmic times is still hampered by 
the fact that most of
the observed {\it Swift} GRBs are without redshift. Indeed,
the measure of the distance has been secured 
for only $\sim 1/3$ of the cases. Given the low completeness level in
redshift determination, the effect of possible observational biases could be
important in shaping their redshift distribution (Fiore et al. 2007). This fact 
strongly limit the possibility of well grounded statistical studies of the 
rest-frame properties of long GRBs and their evolution with cosmic time.
Therefore, it is of paramount importance to obtain an unbiased complete 
sample of GRBs, capable to fully represent this class of object.

To this end, we present in this paper a well selected sub-sample of the full {\it Swift} database.
We select bursts that have favorable observing conditions for redshift
determination from ground  and that are bright in the 15-150 keV {\it Swift}/BAT band.
We find 58 bursts matching our selection criteria
with a completeness level in redshift determination of $90$\%. The completeness
level increases up to $\sim 95$\% by considering the redshift constraints imposed by
the detection of the afterglow or host galaxy in some optical filters.
 Therefore, our selection criteria 
allow us to construct a sizable sample of long bursts that is (almost) 
complete in redshift, providing
the solid basis for the  
study of the long GRB population in an unbiased way. 
In particular, since our selection is based on the brightness in the
{\it Swift}/BAT band, our sample is not biased against the detection
of dark bursts, thus providing a complete description of the whole
long GRB population. 

In the present paper, we will take advantage of the high completeness level of 
our sample to constrain the GRB luminosity function (LF) and its evolution 
with cosmic time. In the last few years, this problem has been faced by many 
different authors ((e.g. Porciani \& Madau 2001, Firmani et al. 2004, 
Guetta et al. 2005, Natarajan et al. 2005, Daigne et al. 2006, 
Salvaterra \& Chincarini 2007, Salvaterra et al. 2009b, Butler et al. 2010, 
Wanderman \& Piran 2010, Campisi et al. 2010, Qin et al. 2010, 
Virgili et al. 2011, Robertson \& Ellis 2012). There is a general agreement about
the fact that GRBs must have experienced some sort of evolution through 
cosmic time, whereas the nature and the level of such evolution is still
matter of debate. Most of the previous works relied on the 
assumption that bursts lacking of redshift measurements follow closely the 
redshift distribution of bursts with known $z$. 
In the past, we tried to overcome
this assumption by deriving conservative lower limit on the level of 
evolution on the basis of the number of bursts detected at $z>2.5$ 
(Salvaterra \& Chincarini 2007) and of bursts with peak luminosity 
$L>10^{53}$ erg s$^{-1}$ (Salvaterra et al. 2009b). 
For the first time, thanks to our well selected, complete 
sub-sample of {\it Swift} GRBs, we can tackle this issue in a unbiased way.
In future works we will use this sample to study
the correlation between physical
parameters of the bursts, the properties of the burst light
curves and of the environment in which they explode.

This paper is organized as follow. In Sect.~2 we describe our selection 
criteria and present our sample. We present our models of the GRB LF and
redshift distribution in Sect.~3, while the results are given in Sect.~4.
We extrapolate our findings to the detection limit of {\it Swift} in Sect.~5.
Finally, in Sect.~6 we draw our conclusions.

\begin{table*}
\begin{center}
\begin{tabular}{lccclccclccclcc}
\hline
\hline
GRB & redshift & ref. & & GRB & redshift & ref & & GRB & redshift & ref & & GRB & redshift & ref \\
\hline
                         
050318 & 1.44 & 1 &  $|$ & 061007 & 1.26 & 2  & $|$  &080603B & 2.69 & 2   & $|$ & 090709A & $<3.5$ & 15	     \\ 
050401 & 2.90 & 2  & $|$ &  061021 & 0.35 & 2 &  $|$  &  080605 & 1.64 & 2 & $|$ & 090715B & 3.00 & 16 \\
050416A & 0.65 & 3 & $|$ & 061121 & 1.31 & 2 &  $|$ &  080607 & 3.04 & 2   & $|$ & 090812 & 2.45 & 17 \\
050525A & 0.61 & 4 & $|$ & 061222A & 2.09 & 3 & $|$ &   080613B & - & -   & $|$ & 090926B & 1.24 & 18 \\
050802 & 1.71 & 2 &  $|$ & 070306 & 1.50 & 2 &  $|$  &  080721 & 2.59 & 2 & $|$ & 091018 & 0.97 & 19 \\
050922C & 2.20 & 2 & $|$ & 070328 & $<4^{(c)}$ &  &  $|$  & 080804 & 2.20 & 2 & $|$ & 091020 & 1.71 & 20 \\
060206 & 4.05 & 2 &  $|$ & 070521 & 1.35 & 3 &  $|$  &  080916A & 0.69 & 2 & $|$ & 091127 & 0.49 & 21 \\
060210 & 3.91 & 2 &  $|$ & 071020 & 2.15 & 2 &  $|$ &   081007 & 0.53 & 8  & $|$ & 091208B & 1.06 & 22 \\
060306 & 3.5$^{(a)}$  &  & $|$ & 071112C & 0.82 & 2 & $|$  &  081121 & 2.51 & 9 & $|$ & 100615A & - & - \\
060614 & 0.13 & 2 &  $|$ & 071117 & 1.33 & 2 &  $|$  &  081203A & 2.10 & 10  & $|$ &   100621A & 0.54 & 23 \\
060814 & 1.92$^{(b)}$ &  &  $|$ &080319B & 0.94 & 2 & $|$  & 081221 & 2.26$^{(a)}$ &  & $|$ &       100728B & 2.106 & 24 \\ 
060904A & - & - &    $|$ & 080319C & 1.95 & 2 & $|$  & 081222 & 2.77 & 11  & $|$ &   110205A & 2.22 & 25 \\ 
060908 & 1.88 & 2 &  $|$ & 080413B & 1.1 & 2 & $|$  &  090102 & 1.55 & 12  & $|$ &  110503A & 1.613 & 26,27 \\
060912A & 0.94 & 5 & $|$  & 080430 & 0.77 & 6 & $|$  &  090201 & $<4$ & 13 & $|$ & & &  \\
060927 & 5.47 & 2 &  $|$ & 080602 & $\sim 1.4^{(d)}$ & 7 & $|$ &090424 & 0.54 & 14   & $|$ & & &  \\
\hline
\hline
\end{tabular}
\end{center}
\caption{List of the bursts matching our selection criteria.  Redshifts or 
limits are provided in the following references:
[1] Berger et al. 2005, [2] Fynbo et al. 2009a and references therein, 
[3] Perley et al. 2009, [4] Foley et al. 2005, 
[5] Levan et al. 2007, [6] Cucchiara \& Fox 2008, [7] Rossi et al. 2012, 
[8] Berger et al. 2008, [9] Berger \& Rauch 2008, [10] Kuin et al. 2009, 
[11] Cucchiara  et al. 2008, [12] de Ugarte Postigo  et al. 2009a, 
[13] D'Avanzo et al. 2009, [14] Chornock  et al. 2009, 
[15] Perley et al. in prep., 
[16] Wiersema  et al. 2009a, [17] de Ugarte Postigo  et al. 2009b, 
[18] Fynbo  et al. 2009b, [19] Chen et al. 2009, [20] Xu et al. 2009, 
[21] Cucchiara et al. 2009, [22] Wiersema et al. 2009b, [23] Milvang-Jensen  
et al. 2010, [24] Flores  et al. 2010, [25] Cucchiara  et al. 2011b, 
[26] de Ugarte Postigo  et al. 2011, [27] D'Avanzo et al. 2011. 
}
\begin{flushleft}
\begin{small}

 (a) based on VLT/X-shooter spectra of the host galaxies obtained within the program 087.A-0451 (PI: H. Flores). The spectra were reduced using the X-shooter data reduction pipeline version 1.3.7 (see Goldoni et al. 2006). In the NIR arm spectrum of GRB\,060306 we identified at the afterglow position the [OII] doublet emission at $z=3.5$, while in the NIR arm spectrum of GRB\,081221 the [OIII] and H$\alpha$ emission lines are present at $z=2.26$ at the afterglow position.

(b) for this GRB a redshift of $z=0.84$ was reported by Thoene et al, (2007).
Images and spectra at the afterglow position have been taken also using the
VLT/FORS within the program 177.A-0591 (PI: J. Hjorth). We downloaded these
data from the ESO Archive and reduced them with standard procedures using
the ESO-MIDAS package. At the afterglow position we could identify an
object, showing a continuum signature in the spectra, that we therefore
consider to be the host galaxy of GRB\,060814. We can also identify the
galaxy reported by Thoene et al. (2007), at $z=0.84$, but this object is offset
from the afterglow position. VLT/X-shooter spectroscopy of the host galaxy
has been performed within the program 084.A-0303 (PI: J. Fynbo). We reduced
these spectra using the X-shooter data reduction pipeline version 1.2.0
(see Goldoni et al. 2006). Thanks to the identification in the NIR arm of the [OII]
doublet, [OIII]\,$\lambda$5007 and H$\alpha$ emission lines associated with
the host galaxy, we can establish a redshift of $z=1.92$ for GRB\,060814.

(c) based on the $R-$band host galaxy detection in ESO-VLT/FORS2 imaging data
obtained  within the program 177.A-0591 (PI: J. Hjorth), taken from the
ESO Archive.

(d) photometric redshift on the bases of the most probable host galaxy association 
in the XRT error circle (Rossi et al. 2012).
\end{small}
\end{flushleft}
\end{table*}

\section{The sample}

About $1/3$ of all GRBs observed by the {\it Swift} satellite (Gehrels et
al. 2004) has a measured redshift. While this represents an enormous improvement
with respect to the pre-{\it Swift} situation, the sample is still far
to be considered as complete. Jakobsson et 
al. (2006) proposed a series of criteria in order to carefully select long GRBs 
which have observing conditions favorable for redshift determination. In 
particular, they required that: i) the burst has been well localized by 
{\it Swift}/XRT and its coordinate quickly distributed; ii) the Galactic
extinction in the burst direction is low ($A_V< 0.5$); iii) the GRB 
declination is $-70^\circ<\delta<70^\circ$; iv) the Sun-to-field distance is 
$\theta_{\rm Sun}> 55^\circ$; v) no nearby bright stars is present. While 
none of the above criteria is expected to alter significantly the redshift 
distribution of observed GRBs, the completness level is increased to 
$\sim 53$\%\footnote{Up to May 2011, the sample consists in 248 long GRBs, 
132 with measured redshift. See http://www.raunvis.hi.is/$\sim$pja/GRBsample.html}.
Still this level of completeness is way too low to permit robust population studies.

In order to construct a more complete sample
we restrict ourself to GRBs that are relatively bright in the 15-150 keV 
{\it Swift}/BAT band. In particular, we select bursts matching the above 
criteria and having 1-s peak photon flux $P\ge 2.6$ ph s$^{-1}$ cm$^{-2}$. This
corresponds to an instrument that is $\sim 6$ times less sensitive than
{\it Swift}. 58 GRBs match our
selection criteria and are listed in Table~1 up to May 2011. 52 of them have measured redshift
so that our completeness level is $90$\%. Of these 52, all but two (namely GRB~070521; 
Perley et al. 2009 and GRB~080602; Rossi et al. 2012) have spectroscopic confirmed redshift either from 
absorption lines over-imposed on the GRB optical afterglow or from emission 
lines of the GRB host galaxy. Moreover, for 3 of the 6 bursts lacking measured z
the afterglow or the host galaxy have been detected in at least one optical filter, 
so that $\sim 95$\% of 
the bursts in our sample have a constrained redshift. 
We note that, while our sample represents only $\sim 10$\% of the full {\it 
Swift} sample,  it contains more than 30\% of long GRBs with known redshift.

The redshift distribution of the bursts in our sample is shown in Fig.~\ref{fig:res}.
In spite of the rather severe cut in the
observed photon flux, the bursts in our sample have a broad distribution in 
redshift. The mean (median) redshift of the sample is 
$1.84\pm 0.16$ ($1.64\pm 0.10$) with a long tail at high-$z$
extending, at least, up to $z=5.47$. 

\section{Model description}

The expected redshift distribution of GRBs can be computed once the 
GRB LF and the GRB formation history has been specified. 
We briefly recap here the adopted formalism and  refer the
interested reader to Salvaterra \& Chincarini (2007) and Salvaterra et al. (2009b) 
for the model details.

The observed peak photon flux, $P$, in the energy band 
$E_{\rm min}<E<E_{\rm max}$, emitted by an isotropically radiating source 
at redshift $z$ is

\begin{equation}
P=\frac{(1+z)\int^{(1+z)E_{\rm max}}_{(1+z)E_{\rm min}} S(E) dE}{4\pi d_L^2(z)},
\end{equation}

\noindent
where $S(E)$ is the differential rest--frame photon luminosity of the source, 
and $d_L(z)$ is the luminosity distance. 
To describe the typical burst spectrum we adopt a Band function with low- and 
high-energy spectral index -1 and -2.25, respectively (Band et al. 
1993; Preece et al. 2000; Kaneko et al. 2006).
The spectrum normalisation is obtained by imposing that the 
isotropic--equivalent peak luminosity is $L=\int^{10000\,\rm{keV}}_{1\,\rm{keV}} E S(E)dE$.
In order to broadly estimate the peak energy of the spectrum, $E_p$, 
for a given $L$, we assumed
the validity of the correlation between $E_p$ and $L$ (Yonetoku et al. 2004; 
Ghirlanda et al. 2005, Nava et al. 2011).

Given a normalised GRB LF, $\phi(L)$, the observed rate of 
bursts with peak flux between $P_1$ and $P_2$ is

\begin{eqnarray}
\frac{dN}{dt}(P_1<P<P_2)&=&\int_0^{\infty} dz \frac{dV(z)}{dz}
\frac{\Delta \Omega_s}{4\pi} \frac{\Psi_{\rm GRB}(z)}{1+z} \nonumber \\
& \times & \int^{L(P_2,z)}_{L(P_1,z)} dL^\prime \phi(L^\prime),
\end{eqnarray}

\noindent
where $dV(z)/dz=4\pi c d_L^2(z)/[H(z)(1+z)^2]$ is the comoving volume 
element\footnote{We adopted the `concordance' model values for the
cosmological parameters: $h=0.7$, $\Omega_m=0.3$, and $\Omega_\Lambda=0.7$.},
and $H(z)=H_0 [\Omega_M (1+z)^3+\Omega_\Lambda+(1-\Omega_M-\Omega_\Lambda)(1+z)^2]^{1/2}$.
$\Delta \Omega_s$ is the solid angle covered on the sky by the survey,
and the factor $(1+z)^{-1}$ accounts for cosmological time dilation.
Finally, $\Psi_{\rm GRB}(z)$ is the comoving burst formation rate.

We explore two general expression for the GRB LF: a single power-law with an
exponential cut-off at low luminosity (exponential LF) and a broken
power-law LF. The former is described by:

\begin{equation}\label{eq:LF}
\phi(L) \propto \left(\frac{L}{L_{\rm cut}}\right)^{-\xi_b} \exp \left(-\frac{L_{\rm cut}}{L}\right),
\end{equation}

\noindent
and the latter by:

\begin{equation}\label{eq:LFd}
\phi(L)\propto 
\left\{
\begin{array}{rcl}
\left(\frac{L}{L_{\rm cut}}\right)^{-\xi_f} & {\rm for } & L\le L_{\rm cut} \\
\left(\frac{L}{L_{\rm cut}}\right)^{-\xi_b} & {\rm for } & L> L_{\rm cut}, \\ 
\end{array}
\right.
\end{equation}

\noindent
where $L_{\rm cut}$ is the cut-off (break) luminosity and $\xi_b$ and $\xi_f$ 
are the bright- and faint-end power-law index, respectively. The GRB LF are
normalized to unity. In order to prevent the integral to diverge we adopt
a minimum GRB luminosity $L_{\rm min}=10^{49}$ erg s$^{-1}$. We test that
our results do not change if minimum luminosity of $10^{48}$ erg
s$^{-1}$ is adopted (apart from the value of the normalization $\eta_0$).

\begin{table*}
\begin{center}
\begin{tabular}{lcccccccc}
\hline
\hline
Model & evo. par. & $\eta_0$ &  $L_{0,51}$ & $\xi_f$ & $\xi_b$ & C-stat & AIC \\
\hline
\multicolumn{8}{c}{cut-off luminosity function}\\
\hline 
no evolution & - &  $0.30$ & $1.0^{+0.9}_{-0.5}$ & - & $2.03^{+0.16}_{-0.12}$ & 93 & 99 \\ 
luminosity & $\delta_l=2.3\pm0.6$  & $0.14$ &
$0.22_{-0.13}^{+0.27}$ & - &$2.02_{-0.10}^{+0.13}$ & 27 & 35\\
density & $\delta_n=1.6\pm 0.4$ & $0.03$ & $3.07_{-1.94}^{+3.29}$ & - & $2.09_{-0.17}^{+0.23}$ &  38  & 46 \\
metal & $Z_{\rm th}=0.14\pm 0.16$ & $0.04$ & $4.4_{-2.8}^{+5.6}$ & - & $2.19_{-0.20}^{+0.30}$ & 37 & 45 \\
\hline
\multicolumn{8}{c}{double power-law luminosity function}\\
\hline 
no evolution & - & $2.06$ &  $25_{-21}^{+68}$ & $1.56_{-0.42}^{+0.11}$ & $2.31_{-0.31}^{+0.35}$ & 88 & 96 \\ 
luminosity & $\delta_l=2.1\pm 0.6$ & $0.21$ & $0.55_{-0.34}^{+0.69}$ & $0.74_{-1.36}^{+1.42}$ & $1.92^{+0.14}_{-0.11}$ & 33 & 43\\
density & $\delta_n=1.7\pm 0.5$ & $0.24$ & $38_{-27}^{+63}$ & $1.50_{-0.32}^{+0.16}$ & $2.32_{-0.32}^{+0.77}$ & 27 & 37 \\
metal & $Z_{\rm th}=0.10\pm 0.18$& $0.41$ & $47_{-35}^{+75}$ & $1.45_{-0.35}^{+0.17}$& $2.58_{-0.50}^{+0.60}$ & 26 & 36 \\
\hline
\hline
\end{tabular}
\end{center}
\caption{Best fit parameters for different models. Errors show the 1-$\sigma$ confidence level for the parameters of interest (see text in section 4 for the details). 
 The last two column report the total C-stat
value (i.e. the
sum of the C-stat values obtained from the fit of the BATSE and {\it
  Swift} dataset) and the Akaike Information Criterion (AIC)  score,
respectively.  We note that in
order to properly compare different models the AIC criterion has to be
considered, where $\exp(({\rm AIC}_{\rm min}-{\rm AIC}_i)/2)$ can be interpreted
as the relative probability that the $i$-th model minimizes the
(estimated) information loss with respect  to the model with the
minimum AIC, AIC$_{\rm min}$.
The total number of data points in the fit is 33.
The GRB formation rate at $z=0$,
$\eta_0$, is given in units of Gpc$^{-3}$ yr$^{-1}$ and the characteristic
luminosity $L_{0,51}$ is units of $10^{51}$ erg s$^{-1}$.}
\end{table*}

\section{Model Results}

We optimize the value of the model free parameters, that is the GRB LF, the normalization $\eta_0$ and the evolution parameter, by minimizing the C-stat
function (Cash 1979) jointly fitting the observed differential number counts in 
the 50--300 keV band of BATSE (Stern et al. 2001, 2002) and the observed redshift 
distribution of
bursts in our sample with photon fluxes in excess to 2.6 ph s$^{-1}$
cm$^{-2}$ in the {\it Swift} 15-150 keV band\footnote{For those GRBs lacking of redshift measurement, 
we randomly assign a redshift from a flat $z$-distribution (taking into account
the available redshift constraints) not to introduce any
a-priori bias. We produced 1000 of such random realizations.}.
For BATSE, we adopt 9.1 yrs of observation with an average exposure
factor of 0.47, including both Earth-blocking and average duty cycle
for useful 1.024s continuos record (Stern et al. 2002). While
our complete {\it Swift} sample provides a powerful test for the existence
and the level of evolution of the long GRB population with redshift, the fit
to the BATSE number counts allows us to obtain the present day GRB rate
density and to better constrain the GRB LF free parameters.
It is worth to note that the best-fit parameters provide a good fit also
of the {\it Swift} differential peak-flux number counts once the 15-150
keV band, the FOV of 1.4sr and the observing lifetime of {\it Swift}
are considered (see also Salvaterra \& Chincarini 2007).
The best fit parameter values together with their 1-$\sigma$
confidence level\footnote{The errors at 1-$\sigma$ confidence level on
the parameters of interest (GRB LF and evolution parameter) 
adopting a C-stat increment of 2.30, 3.53 and 4.72 for 2, 3,
and 4 parameters of interest, respectively.} for different models are provided in Table 2. The
corresponding redshift distributions for bursts with $P > 2.6$ ph
s$^{-1}$ cm$^{-2}$ are shown in Fig. 1. 
We test for each model the two different GRB LF 
parametrizations described in the previous section and we report in Fig.~1 the
one that gives the best result.

\subsection{No-evolution model}

In a first simple (no-evolution) scenario we can assume that long GRBs traces
the cosmic star formation and that their LF is constant in redshift 
($L_{\rm cut}(z)=L_{\rm cut,0}$).
In this case the cosmic GRB formation rate is
$\Psi_{\rm GRB}(z)=\eta_0 \Psi_{\star}(z)$, where $\Psi_{\star}(z)$ is the 
normalized cosmic star formation rate (SFR) and $\eta_0$ is the present-day GRB 
formation rate density in units of Gpc$^{-3}$ yr$^{-1}$. We adopt the 
cosmic SFR recently computed by Li (2008) that extended the previous 
work by Hopkins \& Beacom (2006) to higher redshifts.

\begin{figure}
\center{\includegraphics[scale=0.43]{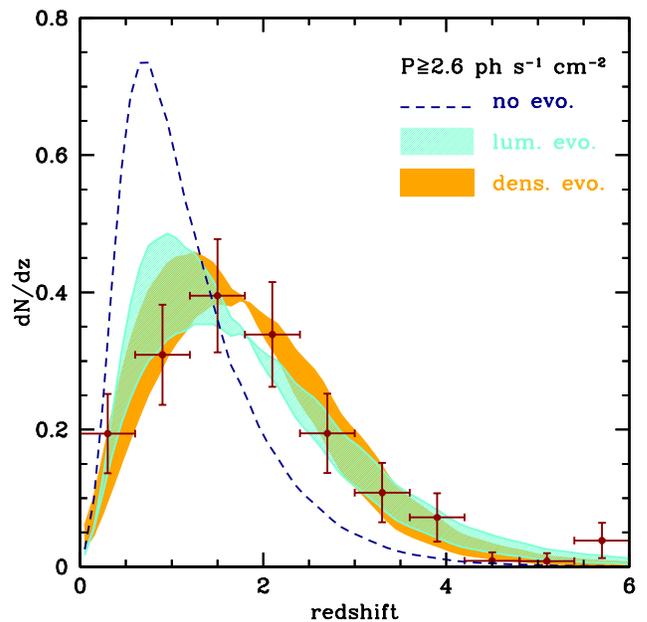}}
\caption{Normalized redshift distribution of GRBs with $P\ge 2.6$ ph s$^{-1}$
cm$^{-2}$. Data points (red) show the observed redshift distribution 
and the error bars show the Poisson uncertainties on the number of detection in
the redshift bin.
The dashed line (blue) shows the expected distribution for the no-evolution case.
Results of luminosity and density evolution models are shown with the
light blue
and dark shaded orange regions, respectively, taking into account the
errors on the evolution parameter. (A color version of this figure is
available in the online journal.)}
\label{fig:res}
\end{figure}

 This no-evolution scenario
(dashed line in Fig.~\ref{fig:res}) clearly does not provide a good representation of the 
observed redshift distribution of our sample,
confirming previous findings (e.g. Daigne et al. 2006, Salvaterra \& Chincarini 2007, 
Salvaterra et al. 2009b, Qin et al. 2010, Wanderman \& Piran 2010, 
Virgili et al. 2011). In particular, the peak
of the GRB redshift distribution is at lower redshift than observed and 
consequently the rate of GRBs at high-$z$ is underpredicted.  This is
confirmed by a more detailed statistical analysis. Indeed, 
on the basis of the Akaike information criterion (Akaike 1974) we can safely discard
this model being $\sim 10^{-14}$ times as probable as the luminosity 
evolution model to minimize the information loss (the density evolution 
model with the broken power-law LF is 0.24 as probable as the luminosity one with
the cut-off LF). Moreover, a KS test between
the no evolution best-fit model and the data of our sample gives a chanche 
probability of $\sim 5\times 10^{-5}$ that the two distribution are drawn from
the same parent population.

In the following sections, we will consider 
evolution scenarios that may enhance the number of detections
at high-$z$. In particular, we explore: i) a luminosity evolution model in which
high-$z$ GRB are typically brighter than low-$z$ bursts and, ii) two density 
evolution models, both leading to an enhancement of the GRB formation rate 
with redshift. 
Hybrid models, with both luminosity and density
evolution, are in principle possible. However, the fit
with hybrid models results to be very degenerate and does not 
provide usefull constraints. Therefore, we prefer here to
consider the two scenarios separately to highlight possible
similarities/differences between the two kinds of evolution.

\subsection{Luminosity evolution model}

Evolution in the GRB LF can provide an enhancement of the high-$z$ GRB
detection, representing  a viable way to 
reconcile model results with the observations.
Here, we consider the possibility that the cut-off (break) luminosity 
is an increasing function of the redshift, that is
$L_{\rm cut}(z)=L_{\rm cut,0}(1+z)^{\delta_l}$. 
We find that a strong luminosity evolution with $\delta_l=2.3\pm 0.6$ 
is required to reproduce the observed redshift distribution of the bursts in
our complete sample (light shaded area in Fig.~\ref{fig:res}). 
The result does not depend on the assumed expression of the GRB
LF. 

\subsection{Density evolution models}

An increase of the rate of GRB formation with redshift (on the top of the known evolution
of the SFR density) will also lead to an enhanced detection of bursts at high-$z$.
As a general case, we parametrize
the evolution in the GRB formation rate as $\eta(z)=\eta_0(1+z)^{\delta_n}$. 
By fitting our datasets we find that strong density evolution is required
with $\delta_n=1.7\pm 0.5$.
The amount of evolution does not depend on the assumed expression of the GBR LF.
However, we note that 
the cut-off LF tends to underestimate the number of low-$z$ 
bursts with respect to the observed one, leading to some discrepancy 
with the first data point.

\begin{figure}
\center{\includegraphics[scale=0.43]{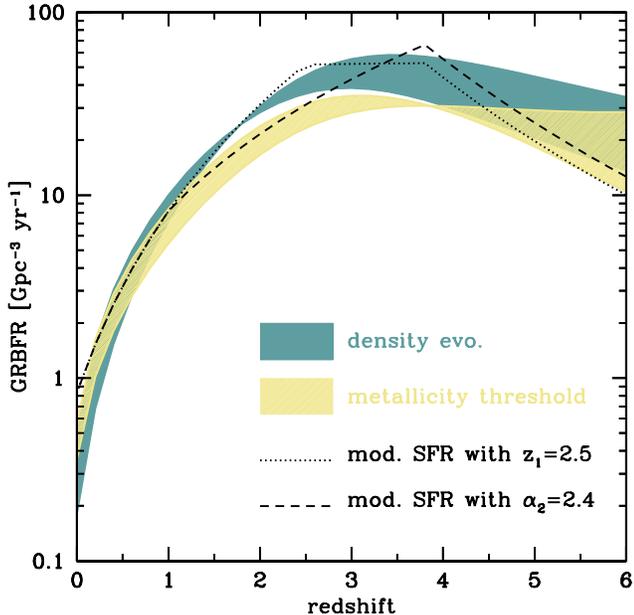}}
\caption{Intrinsic redshift distribution of long GRBs for different density evolution
models. Dark blue and light yellow shaded areas show the results for the density evolution model
and for the metallicity threshold model, respectively. 
Dotted (dashed) lines reports the modified SFR with $z_1=2.5$ ($\alpha_2=2.4$).
In all cases, no evolution of the GRB LF has been assumed. (A color
version of this figure is available in the online journal.)}
\label{fig:dist}
\end{figure}

The large value of $\delta_n$ implies an important shift of the peak of the 
GRB formation rate towards higher redshifts with respect to stars.
We further investigate this issue by applying a ``correction'' to the 
shape of the cosmic SFR. This is usually parametrized as 
three power-laws (Hopkins \& Beacom 2006, Li 2008) with
power-index $\alpha_1=3.3$, $\alpha_2=0.055$, $\alpha_3=-4.46$ and breaks at 
$z_1=0.993$ and $z_2=3.8$ (Li 2008). We fit our datasets by letting one of
the above parameters free to vary in addition to those describing the 
GRB LF. We find that the observed 
redshift distribution of bursts in our sample can be explained either by an 
increase of the redshift of the first break to $z_1=2.5\pm 0.5$ or by a 
hardening of the second power-law to $\alpha_2=2.4\pm 0.4$. 
In both cases, the GRB formation rate is found to peak at a much higher 
redshift with respect to stars. The intrinsic redshift distribution of
GRBs is shown in Fig.~\ref{fig:dist}.

\subsection{Metallicity threshold models}

A subclass of density evolution models foresee the
formation of long GRBs preferentially in low-metallicity environments.
In this case GRBs will be biased tracers of the star formation activity being
their formation suppressed at low-$z$ where most of the galaxies are relatively
metal-rich. Following Langer \& Norman (2006), we model the fractional mass
density belonging to metallicity below a given threshold, $Z_{\rm th}$ as 

\begin{equation}
\Sigma(z)=\frac{\hat{\Gamma}(0.84,(Z_{\rm th}/\Zsun)^2 10^{0.3z})}{\Gamma (0.84)},
\end{equation}

\noindent
where $\hat{\Gamma}$ ($\Gamma$) are the incomplete (complete) gamma
function, and $\Gamma(0.84)\simeq 1.122$. The GRB formation rate is then 
given by $\Psi_{\rm GRB}(z)\propto \Sigma(z)\Psi_\star(z)$.  

We fit our datasets letting $Z_{\rm th}$ free to vary. The 
available data are well described by models with $Z_{\rm th}\le 0.3\;\Zsun$,
almost independently on the assumed LF. The resulting LF is similar to 
the one obtained for density evolution model.
Indeed, the two predicted redshift distributions match each others within
the uncertainties and we refer
to the dark shaded curves in Fig.~\ref{fig:res} also for the metallicity
threshold model.

The range of values for $Z_{\rm th}$ found in our analysis is in agreement 
with the expectation of the collapsar model (Woosley \& Heger 2006; Fryer et 
al. 1999). However, such strong metallicity
cut-offs seem to be inconsistent with the observed properties of GRB host
at $z<1$ (Mannucci et al. 2011; Kocevski et al. 2011; Campisi et at. 2011a).
In particular, Campisi et al. (2011a) have shown that in the presence of a strong
metallicity cut-off for the GRB progenitor star, the expected distribution of
GRB host galaxies in the M-Z and Fundamental Metallicity Relation planes is much
flatter than observed.
Larger metallicity thresholds will 
require some luminosity evolution in order to reproduce the available data. For
$Z_{\rm th}=0.5\;\Zsun$, the typical burst luminosity should increase with redshift
as $(1+z)^{1.3\pm 0.6}$.

\section{{\it Swift} redshift distribution}

\begin{figure}
\center{\includegraphics[scale=0.43]{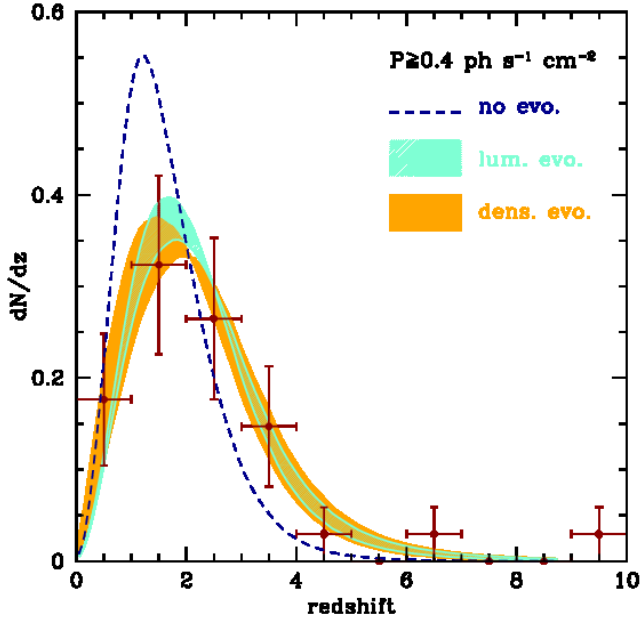}}
\caption{Normalized redshift distribution of GRBs detectable with {\it Swift},
i.e. with observed photon flux in excess to $P_{\rm lim}=0.4$ ph s$^{-1}$
cm$^{-2}$. The data show the observed redshift distribution 
as reported by Greiner et al. (2011), on the basis of the almost complete
sample of GRBs detected by GROND.
Error bars show the Poisson 
uncertainties on the number of objects in the redshift bin.
Model results are shown as in Fig.~\ref{fig:res}. No attempt to fit
the observed redshift distribution has been done. (A color version of this figure is
available in the online journal.)}
\label{fig:swift}
\end{figure}

We compute the redshift distribution expected for the full {\it Swift} dataset
assuming a photon flux limit of $P=0.4$ ph s$^{-1}$ cm$^{-2}$ and fixing
the model free parameters to the values given in Table~2. The results are shown
in Fig.~\ref{fig:swift} for the different evolution scenarios here explored.
The models are compared with the redshift distribution inferred from the sample
of GRBs observed by GROND (Greiner et al. 2011). This sample has a
completeness level similar to our but is smaller in size and cover a  
broader redshift range. We find that our evolution models provide a good 
description of the observed redshift distribution of the GROND sample 
without the need of any adjustment of the free paramters 
(a KS test gives a probability of 50\%), whereas the no evolution model is excluded (probability of $5\times 10^{-4}$). 
This further confirms the reliability
of our analysis strenghtening our conclusions.

The predicted redshift distribution of bursts detectable by {\it Swift} presents
a steep raise at low-$z$ peaking at $z\sim 2$ with a tail extending at 
higher redshifts. The median redshift of the distribution is $z=2.05\pm 0.15$
where the error takes into account the uncertainties on the evolution 
parameters.
We predict that $3-5$\% of the bursts lie at $z>5$, consistently with 
the observational estimate of $5.5\pm 2.8$\% (Greiner et al. 2011). This further
confirms that the majority of dark GRBs are not high-$z$ sources but
more likely obscured by dust (Perley et al. 2009, Greiner et al. 2011). 
At $z>8$, we expect $1.3-3.5$ GRBs\footnote{We do not consider here the
possible contribution of PopIII GRBs that may provide additional GRBs at very
high-$z$ (Campisi et al. 2011b; de Souza et al. 2011).} among the 530 
{\it Swift} GRBs. This is 
consistent with the two detections reported so far, i.e. GRB 090423 at 
$z=8.2$ (Salvaterra et al. 2009; Tanvir et al. 2009) and GRB 090429B at 
$z\sim 9.4$ (Cucchiara et al. 2011). 

It is worth to note that the two evolution scenarios here explored 
predict very similar distributions. Therefore we can not
distinguish between luminosity and density evolution simply
on the basis of the {\it Swift} observed redshift distribution.

\section{Conclusions}

We select a sub-sample of {\it Swift} long GRBs that is
complete in redshift. The sample is composed by bursts with favourable
observing conditions and with  1-s peak photon fluxes
$P\ge 2.6$ ph s$^{-1}$ cm$^{-2}$.
It contains  58 bursts with a completeness level of $\sim 90$\% and provides
the basis for statistical studies of the properties of long GRBs and their evolution
with redshift in a unbiased way.
GRBs can be used to study fundamental issues in astronomy and astrophysics, such as 
the star formation rate and the stellar and metal abundances evolution. They can be use
as tracers of the galaxy evolution, of the ISM composition and to investigate the early universe.
Complete and fully representative samples of GRBs are therefore unique tools to perform these investigations.

Here, we use the observed burst redshift distribution
of our complete sample to probe and constrain the evolution
of the long GRB population in redshift. We confirm that GRBs must have
experienced some sort of evolution being more luminous or more numerous in 
the past than observed today. 
We found that in order to match the observed distribution, the typical burst
luminosity should increase as $(1+z)^{2.3\pm 0.6}$ or the GRB rate density 
as $(1+z)^{1.7\pm 0.5}$ on the top of the known cosmic evolution of the SFR. 
This result does not depend on the assumed expression of
the GRB luminosity function. We also explore models in which GRBs form 
preferentially in low-metallicity environments. We find
that the metallicity threshold for GRB formation should be lower than 0.3 
$\Zsun$ in order to account for the observations assuming no evolution of
the GRB LF. This value, while consistent
with the expectations of collapsar models, seems to be at odd with the 
observed properties of $z<1$ GRB hosts  (Levesque et al. 2010;
Mannucci et al. 2011; Campisi et al 2011a). 

Extrapolating our results to $P=0.4$ ph s$^{-1}$ cm$^{-2}$,
the predictions of evolution models are consistent with available
observational constrain without the need of any adjustment of the
model free parameters. We also note that the predicted redshift distributions
at the sensitivity of {\it Swift} for different evolution models are very similar, 
implying that it would be extremely difficult to distinguish among different kind of 
evolutions only on the basis
of the observed redshift distribution. 

On the bases of our models we predict that $3-5$\% of the bursts detected by
{\it Swift} lie at $z>5$ consistently with the most recent observational estimate
of $5.5\pm 2.8$\% (Greiner et al. 2011). This indicates that high-$z$ bursts can
contribute only marginally to the observed fraction of dark bursts. 
Finally, we expect 30-50 bursts per year over the entire sky with luminosities
in excess to $10^{53}$ erg s$^{-1}$ and 2-5 with luminosities exceeding $10^{54}$ erg s$^{-1}$. 
Extreme luminous GRBs, such as the current {\it Swift}
record holder GRB~080607 with $L_{iso}= 2.26\times 10^{54}$ erg s$^{-1}$ (Perley et al. 2011), should 
be rare, exploding once every 0.5-1 years in the Universe. This corresponds to one detection 
every $\sim 0.7-1.5$ years in the field of view of {\it Fermi}/GBM (and one every 5-12 years for 
{\it Swift}). 

\section*{Acknowledgments}

We thank A. Rossi and D. Perley for sharing their results before publication. 
This work has been supported by ASI grant I/004/11/0.

\end{document}